\documentclass[aps,prl,twocolumn,amsmath,amssymb,superscriptaddress]{revtex4}%
\usepackage{graphicx}
\usepackage{dcolumn}
\usepackage{bm}
\usepackage{color}
\usepackage{amsmath}
\usepackage{epstopdf}
\usepackage{fancybox}
\usepackage{amsfonts}
\usepackage{amssymb}%
\usepackage[toc,page]{appendix}
\renewcommand{\cite}[1]{{[}\onlinecite{#1}{]}}

\newcommand{\pa}{\partial}

\newcommand{\be}{\begin{equation}}
\newcommand{\e}{\end{equation}}
\newcommand{\beml}{\begin{subequations}}
\newcommand{\eml}{\end{subequations}}
\newcommand{\beq}{\begin{eqnarray}}
\newcommand{\eq}{\end{eqnarray}}
\newcommand{\ba}{\begin{array}}
\newcommand{\ea}{\end{array}}
\newcommand{\bpm}{\begin{pmatrix}}
\newcommand{\epm}{\end{pmatrix}}
\newcommand{\bc}{\begin{cases}}
\newcommand{\ec}{\end{cases}}
\newcommand{\lt}{\left}
\newcommand{\rt}{\right}

\newcommand{\bb}{\boldsymbol}

\begin{document}
\title{Theory of the Interfacial Dzyaloshinskii-Moriya Interaction in Rashba Antiferromagnets}

\author{Alireza Qaiumzadeh}
\affiliation{Center for Quantum Spintronics, Department of Physics, Norwegian University of Science and Technology, NO-7491 Trondheim, Norway}

\author{Ivan A. Ado}
\affiliation{Radboud University, Institute for Molecules and Materials, 6525 AJ Nijmegen, Netherlands}

\author{Rembert A. Duine}
\affiliation{Center for Quantum Spintronics, Department of Physics, Norwegian University of Science and Technology, NO-7491 Trondheim, Norway}
\affiliation{Institute for Theoretical Physics and Center for Extreme Matter and Emergent Phenomena,
Utrecht University, Princetonplein 5, 3584 CC Utrecht, Netherlands}
\affiliation{Department of Applied Physics, Eindhoven University of Technology,
P.O. Box 513, 5600 MB Eindhoven, Netherlands}

\author{Mikhail Titov}
\affiliation{Radboud University, Institute for Molecules and Materials, 6525 AJ Nijmegen, Netherlands}
\affiliation{ITMO University, Saint Petersburg 197101, Russia}

\author{Arne Brataas}
\affiliation{Center for Quantum Spintronics, Department of Physics, Norwegian University of Science and Technology, NO-7491 Trondheim, Norway}

\begin{abstract}
In antiferromagnetic (AFM) thin films, broken inversion symmetry or coupling to adjacent heavy metals can induce Dzyaloshinskii-Moriya (DM) interactions. Knowledge of the DM parameters is essential for understanding and designing exotic spin structures, such as hedgehog Skyrmions and chiral N\'{e}el walls, which are attractive for use in novel information storage technologies. We introduce a framework for computing the DM interaction in two-dimensional Rashba antiferromagnets. Unlike in Rashba ferromagnets, the DM interaction is not suppressed even at low temperatures. The material parameters control both the strength and sign of the interfacial DM interaction. Our results suggest a route toward controlling the DM interaction in AFM materials by means of doping and electric fields.
\end{abstract}

\date{\today}
\maketitle

Relativistic spin-orbit coupling (SOC) is the foundation of spin orbitronics, a rapidly developing branch of spintronics \cite{Manchon-Rashba1, Fert-Rashba3}. Anisotropic magnetoresistance \cite{Thomson} and the anomalous Hall effect \cite{Hall} are established SOC transport phenomena. More recent discoveries include the spin Hall and inverse spin Hall effects \cite{SHE}, topological surface states \cite{Fert-Rashba3,TSS1,TSS2}, spin-orbit torques \cite{Manchon-Rashba1, ManchonSOT, Alireza-Rashba2, AFM-Hamiltonian1, AFM-Hamiltonian2}, and chiral domain walls and Skyrmions \cite{skyrmion2009,Hoffmann,Emori,Thiaville,Alireza-skyrmion,Alireza-AFM1,Yudin}. These phenomena are essential to enable novel ultrafast, nonvolatile, nanoscale spin-based storage and computation devices.

\begin{figure}[t]
\includegraphics[width=\columnwidth]{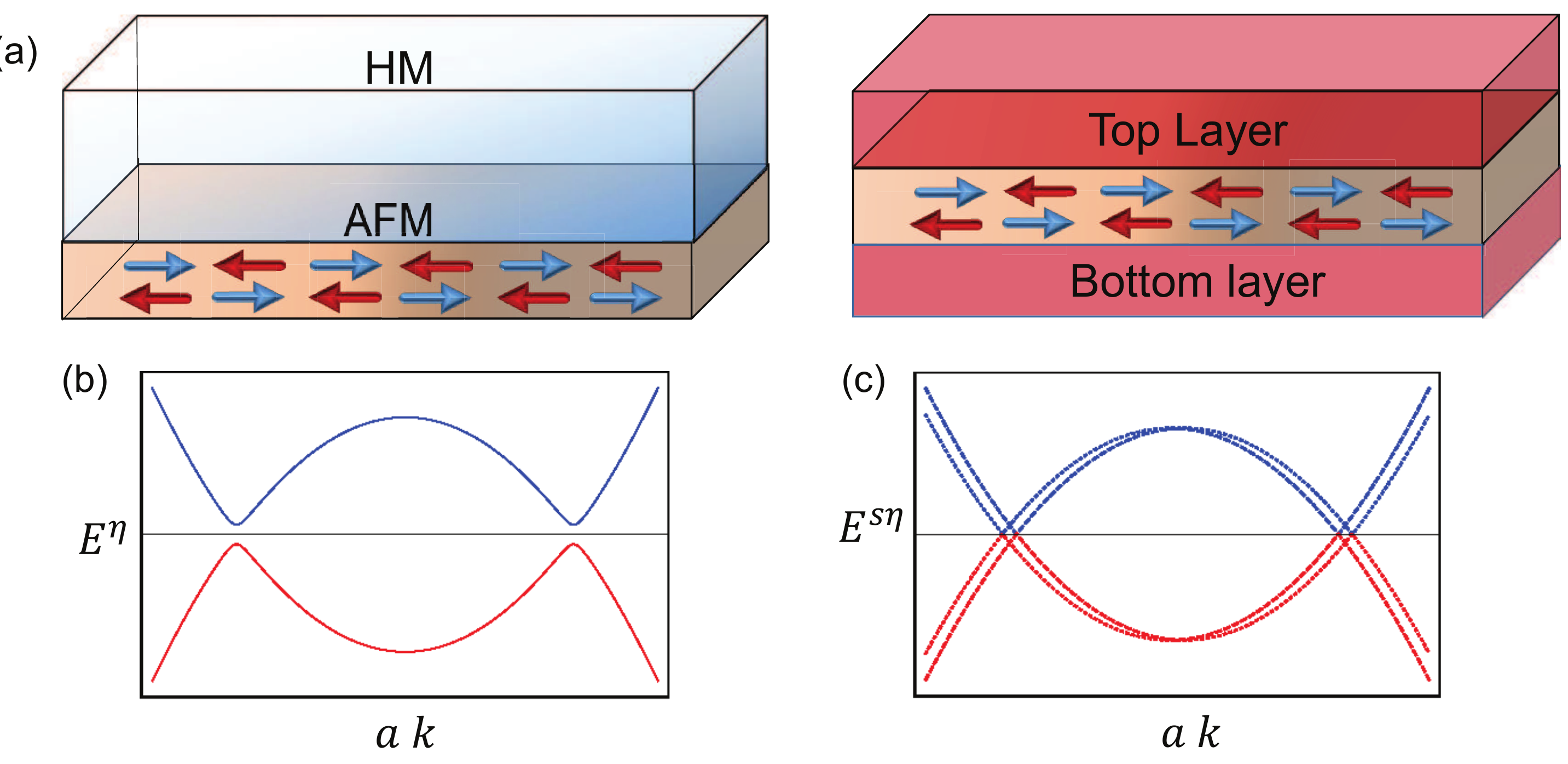}
\caption{(a) The Rashba-AFM model is a model for systems with inversion asymmetry: an AFM/HM bilayer (left) and an AFM thin film asymmetrically embedded between two different insulator layers (right). (b) , (c) Electronic dispersion relations in the limit of strong exchange coupling (b) and in the limit of high SOC (c). The superscript $\eta=\pm$ refers to the conduction (blue) and valence (red) bands, whereas the superscript $s=\pm$ specifies chiral bands. In these two extreme limits, the band structure is almost isotropic for an AFM system with in-plane anisotropy.}
\label{fig:setup}
\vspace*{-6pt}
\end{figure}

The Dzyaloshinskii-Moriya (DM) interaction between localized spins in a magnetic material is induced by SOC \cite{Dzyaloshinsky, Moriya}. The DM interaction is important for stabilizing noncollinear magnetic structures in ferromagnets. The original proposal by Dzyaloshinskii and Moriya, however, concerned antiferromagnetic (AFM) systems with weak magnetization, i.e., weak ferromagnets; this weak magnetism can be explained in terms of antisymmetric exchange, which is also referred to as the DM interaction.

The energy associated with the DM interaction between two neighboring spins in a lattice, $\bm{S}_{1}$ and $\bm{S}_{2}$, can be written in the form of the mixed product $\mathcal{H}_{DM}=-\bm{D}_{12} \cdot \bm{S}_1 \times \bm{S}_2$, where $\bm{D}_{12}$ is called the DM vector. The DM interaction, therefore, favors a perpendicular orientation of neighboring spins. By contrast, the Heisenberg exchange interaction $\mathcal{H}_{\mathrm{ex}}=J_{\mathrm{ex}} \bm{S}_1 \cdot \bm{S}_2$ favors a collinear magnetic order, which can be either ferromagnetic (FM) for $J_{\mathrm{ex}}<0$ or AFM for $J_{\mathrm{ex}}>0$. The competition between the Heisenberg exchange interaction and the DM interaction leads to the formation of exotic structures such as chiral domain walls, helices, and Skyrmions \cite{Fert-Rashba3}.

In the bulk of a noncentrosymmetric magnetic crystal, the DM vector points along one of the lattice vectors and gives rise to so-called Bloch-like structures \cite{Dzyaloshinsky, Moriya, iDMI5}. In an ultrathin magnetic film with impurity- or interface-induced SOC, there is also an interfacial DM interaction that corresponds to a DM vector pointing out along the interface. This gives rise to so-called N\'{e}el-type textures \cite{iDMI1,iDMI2,iDMI3,iDMI4,iDMI5,Thiaville, iDMI7}. Both the sign and the amplitude of the DM vector are of critical importance for observing and engineering different chiral structures \cite{iDMI-exp0,katsnelson,iDMI-exp1,iDMI-exp2,iDMI-exp3,iDMI-exp4,iDMI-exp5,iDMI-exp6,iDMI-exp7,iDMI-exp8,iDMI-exp9,iDMI-exp10,iDMI-exp11,iDMI-exp12, AFMDMI-exp,iDMI-abinitio1,iDMI-abinitio2}.

AFM materials have recently attracted considerable attention as active spintronics elements \cite{AFM-review,AFM-review1}. The absence of stray fields, the possibility of operating at terahertz frequencies, and the existence of spin waves with opposite helicities make antiferromagnets promising candidates for use in the next generation of spin-based memory and processing devices \cite{AFM-review,AFM-review1} with a nanoscale element base.

The coexistence of strong SOC and spin ordering at the interfaces of AFM/heavy-metal (HM) bilayers makes such heterostructures particularly promising for low-dimensional spin orbitronics applications \cite{Interface-review}. An AFM thin film sandwiched between insulators is also a common functional geometry for spin-orbitronics due to interfacial SOC. The interfacial SOC in such a system is effectively described by Rashba SOC \cite{Manchon-Rashba2}. For magnetic films, the two-dimensional (2D) Rashba model captures the main physics and trends of SOC with a broken inversion symmetry \cite{Manchon-Rashba2}, such as fieldlike and dampinglike spin-orbit torques \cite{Manchon-Rashba1, Alireza-Rashba2, AFM-Hamiltonian1}, intrinsic spin Hall effects \cite{SHE}, intrinsic anomalous Hall effects \cite{AHE}, inverse Faraday effects \cite{Alireza-Rashba1}, and magnetic anisotropy \cite{Rashba-anisotropy}.

In this Letter, we develop a framework for computing the interfacial DM interaction in AFM layers with inversion asymmetry. The typical system illustrated in Fig.~\ref{fig:setup} is described on the basis of an effective 2D AFM-Rashba Hamiltonian. We find that both the sign and the magnitude of the DM vector depend on the ratio relating three energy scales: the chemical potential, the \textit{s-d} exchange interaction, and the Rashba SOC strength. In particular, the strong dependence on the chemical potential suggests that the DM interaction can be tuned by modifying the electron density by means of doping or voltage gating \cite{Fechner2012}.

A generic effective 2D Hamiltonian describing itinerant electrons in the AFM layer [see Fig.~\ref{fig:setup}(a)] can be written as
\begin{align}
\mathcal{H}=\mathcal{H}_{\mathrm{kin}}+\mathcal{H}_{\mathrm{sd}}+\mathcal{H}_{\mathrm{so}}, \label{hamiltonian}
\end{align}
where $\mathcal{H}_{\mathrm{kin}}$ is the kinetic energy of the electrons, $\mathcal{H}_{\mathrm{sd}}=J_{\mathrm{sd}}\,\bm{n} \cdot \bm{\Gamma}$ describes an effective interaction with a strength $J_{\mathrm{sd}}$ between the spins of the itinerant $s$ electrons and the localized $d$ electrons \cite{sd-model-referee-suggested}, and $\mathcal{H}_{\mathrm{SO}}$ describes the SOC. The operator $\bm{\Gamma}$ is the direct product of the electron spin operator and the sublattice position operator, which, in the case of an AFM system, accounts for the effects of sublattice staggering. The unit vector $\bm{n}$ is the order parameter, which can represent either the total magnetization in a ferromagnet or the staggered magnetization in an antiferromagnet.

We compute the electronic contributions to the DM interaction parameter $D$ and to the exchange stiffness $A$ in the following way. First, we evaluate how the itinerant electrons influence the magnetic subsystem by finding an effective action. We expand the effective action up to linear order with respect to the deviation of the spins from their equilibrium direction. The corresponding susceptibility tensor describes the influence of the electronic degrees of freedom on the localized magnetic moments. A linear expansion of the susceptibility tensor in spatial gradients of $\bb{n}$ defines the DM interaction strength $D$. The contribution of the itinerant electrons to the exchange stiffness $A$ is extracted from the second-order expansion in spatial gradients.

The action $\mathcal{S}$ defines the system partition function $\mathcal{Z}=\int d[\Phi^*] d[\Phi] d[\bm{n}] e^{-\mathcal{S}[\Phi^*,\Phi,\bm{n}]/\hbar}$, where $\Phi$ is the Grassmannian coherent-state spinor and $\hbar$ is Planck's constant. In the $s$-$d$ approach to magnetic systems, the action is decomposed into the sum $\mathcal{S}=\mathcal{S}_F+\mathcal{S}_B$, where $\mathcal{S}_F[\Phi^*,\Phi,\bm{n}]$ is the fermionic action corresponding to the Hamiltonian of Eq.~(\ref{hamiltonian}), which also includes the $s$-$d$ coupling, and $\mathcal{S}_B[\bm{n}]$ is the bosonic action describing the dynamics of the localized spins (magnons) in the absence of itinerant electrons. In our model, it is the coupling between the itinerant electrons and the local moments that determine the DM interaction, which is also directly linked to the SOC of the itinerant electrons. We will not specify the bosonic part of the action $\mathcal{S}_B[\bm{n}]$ since it is irrelevant for the subsequent discussion.

The fermionic action reads
\be
\mathcal{S}_F\!=\!\int\!\!\!\int_0^{\hbar\beta}\!\!d\tau d\tau'\!\! \int\!\!\!\int\! d{\bm{r}} d{\bm{r}}' \Phi^*_{\bm{r},\tau} [-\hbar  G^{-1}_{\bm{r},\tau;\bm{r}',\tau'}] \Phi_{\bm{r}',\tau'},
\e
where $\beta=1/k_B T$ is the inverse temperature and $\tau$ is the imaginary time. The inverse Green's function operator is  $\hbar\, G^{-1}_{\bm{r},\tau;\bm{r}',\tau'}=-\left(\hbar{\partial_\tau}+\mathcal{H}\right)\delta(\bm{r}-\bm{r}')\delta(\tau-\tau')$ in terms of the Hamiltonian of Eq.~(\ref{hamiltonian}). We compute the effective theory for the vector field $\bb{n}(\bb{r})$ by integrating out the fermionic degrees of freedom. This standard procedure results in an additional, effective contribution to the bosonic action of the form $\Delta \mathcal{S}_F^{\mathrm{eff}}[\bm{n}]=\int_0^{\hbar\beta} d\tau \int d{\bm{r}} \left(-\hbar \mathrm{Tr}[\ln(-G^{-1})]\right)$. Below, we analyze $\Delta \mathcal{S}_F^{\mathrm{eff}}[\bm{n}]$ and its influence on the magnet in the AFM-Rashba model.

\begin{figure}[t]
\includegraphics[width=\columnwidth]{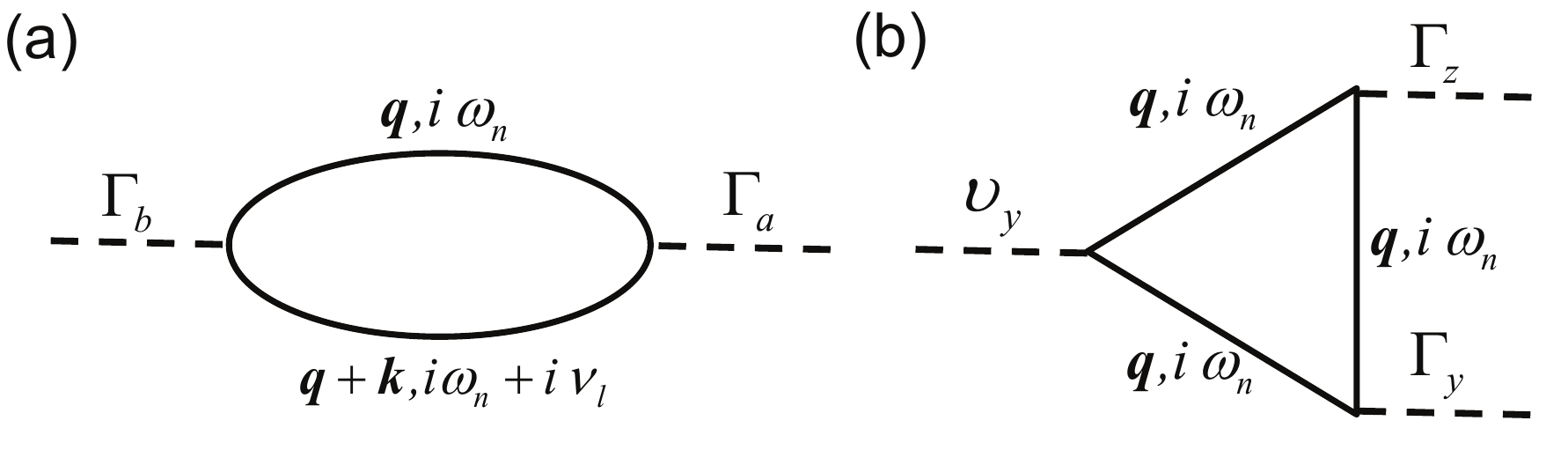}
\caption{ (a) The Feynman diagram corresponding to Eq.~(\ref{polarizibility-pi}) for a single-loop polarizability tensor. (b) The Feynman diagram corresponding to Eq.~(\ref{DMI-formula22}), describing the interfacial DM interaction for a system with inversion symmetry breaking in the $z$ direction and an order parameter vector $\bm{n}$ in the $x$ direction.}
\label{fig:diag}
\end{figure}

In our system, the symmetry-breaking direction ($z$ direction) is perpendicular to the plane \cite{iDMI-exp3}. Without the loss of generality, we choose the $x$ axis to be in the direction of the in-plane vector field $\bb{n}$. A small deviation of the unit vector from its equilibrium direction is, then, parameterized by $\bm{n}=\hat{\bb{x}}+\delta\bm{n}$, where $\delta\bm{n}=\left( -(\delta n^2_y+\delta n^2_z)/2, \delta n_y, \delta n_z \right)$. The effective action is obtained from a perturbation with respect to $\delta \bb{n}$, \cite{sd-model-referee-suggested} that holds irrespective of the value of the exchange strength $J_{sd}$. The effective action is conveniently established from the Dyson equation, $G^{-1}={G^0}^{-1}-\Sigma$, where the unperturbed Green's function refers to $\delta\bb{n}=0$ and the self-energy $\hbar \Sigma=J_{\mathrm{sd}}\delta\bm{n}\cdot\bm{\Gamma}\, \delta(\bm{r}-\bm{r}')\delta(\tau-\tau')$ is introduced.

Computing the self-energy up to the second order in $\delta\bm{n}$ yields \cite{kzero}
\begin{align}
\label{eff-action4}
\Delta \mathcal{S}^{\mathrm{eff}}[\bm{n}]=\hbar\beta\sum_{\bm{k}\neq 0,l}\delta n^a_{\bm{k},\nu_l} \Pi^{ab}_{\bm{k},i\nu_l} \delta n^b_{-\bm{k},-\nu_l},
\end{align}
where the indices $a, b=(y,z)$ denote the transverse vector components with respect to the equilibrium $\hat{\bb{x}}$ direction, $\nu_l=2 n\pi/ \beta$ denotes the bosonic Matsubara frequencies, and
\be
\label{polarizibility-pi}
\Pi^{ab}_{\bm{k},i\nu_l}=\frac{J^2_{\mathrm{sd}}}{2\hbar\beta}\sum_{\bm{q},n}\mathrm{Tr}[\Gamma_a G^0_{\bm{q},i\omega_{n}}\Gamma_b G^0_{\bm{q}+\bm{k},i\omega_{n}+i\nu_l}]
\e
is the dynamical susceptibility tensor, pictured schematically in Fig.~\ref{fig:diag}a. Here, $G^0_{\bm{q},i\omega_{n}}=(i\hbar\omega_n-\mathcal{H})^{-1}$ is the equilibrium Green's function, and the $\omega_n=(2 n+1)\pi/ \beta$ are the fermionic Matsubara frequencies. We compute the sum over the fermionic Matsubara frequencies by using the identity $\sum_n (i \hbar\omega_n-E)^{-1}/\beta=f(E)$, where $f(E)$ is the Fermi distribution.

By expanding the static limit, $\nu_l=0$, on the spin susceptibility to the second order in the wave vector $\bm{k}$, we find both the electronic contribution to the symmetric Heisenberg exchange stiffness, which is determined by the symmetric terms in the diagonal elements of the susceptibility tensor, and the antisymmetric exchange interaction (DM interaction), which is determined by the antisymmetric terms in the off-diagonal elements. From the partition function of the canonical ensemble, $\mathcal{Z}=\int d[{\bm{n}}] e^{-\mathcal{S}^{\mathrm{eff}}[\bm{n}]/\hbar}=\int d[{\bm{n}}] e^{-\beta F[{\bm{n}}]}$, we obtain $\delta_{{\bm{n}}}\mathcal{S}^{\mathrm{eff}}/\hbar =\beta\delta_{\bm{n}} F[{\bm{n}}]$, where the micromagnetic free energy, including the stiffness and the DM interaction, is $F[{\bm{n}}]=\int d^2\bm{r} \left(A (\nabla {\bm{n}})^2-D {\bm{n}}\cdot(\hat{\bb{z}}\times \nabla)\times {\bm{n}}\right)$. By comparing the microscopic free energy with the expression for the effective action, Eq. (\ref{eff-action4}), we define the micromagnetic parameters $A$ and $D$, which characterize the free carrier contributions to the exchange stiffness and the DM interaction, respectively.

Upon expanding the off-diagonal elements of the tensor $\Pi$ to the first order in the wave vector $\bm{k}$ [see Fig.~\ref{fig:diag}(b)], we obtain the relation
\begin{subequations}
\label{DMI-formula22}
\begin{align}
D=&\lt.i\frac{\partial\Pi^{yz}_{\bm{k}}}{\partial {k_y}} \rt|_{\bm{k}=0}=-\lt.i\frac{\partial\Pi^{zy}_{\bm{k}}}{\partial {k_y}} \rt|_{\bm{k}=0}\\
\label{DMI-formula2}
=&i\frac{J^2_{\mathrm{sd}}}{2\beta}\sum_{\bm{q},n}\mathrm{Tr}[\Gamma_y G^0_{\bm{q},i\omega_{n}}\Gamma_z G^0_{\bm{q},i\omega_{n}}v_y G^0_{\bm{q},i\omega_{n}}],
\end{align}
\end{subequations}
where $v_y=\hbar^{-1}\partial \mathcal{H}_{\bm{q}}/\pa q_y$ is the $y$ component of the velocity operator. From the second-order terms, we obtain the electron contribution to the exchange stiffness:
\begin{align}
\label{stiffness}
A=&-\lt.\frac{\partial^2\Pi^{yy}_{\bm{k}}}{\partial {k^2_x}} \rt|_{\bm{k}=0}=-\lt.\frac{\partial^2\Pi^{zz}_{\bm{k}}}{\partial {k^2_y}} \rt|_{\bm{k}=0} \, .
\end{align}
In our model, $A$ describes the contribution to the AFM exchange interaction from a superexchange-type interaction between the localized spins in the AFM layer via the itinerant spins.

We should emphasize here that in this approach, we have ignored the spin fluctuations of the localized AFM spins, which is a valid omission as long as the system temperature is much less than the critical N\'{e}el temperature.

To model an AFM system with interfacial SOC, we use the 2D AFM-Rashba Hamiltonian \cite{AFM-Hamiltonian1,AFM-Hamiltonian2} on a square lattice:
\begin{align}
\label{hamiltonian-AFM}
\mathcal{H}= \gamma_k \tau_x\sigma_0+J_{\mathrm{sd}} \tau_z\bm{\sigma}\cdot\bm{n}-\alpha_R \tau_x(\bm{\sigma}\times\bm{k})\cdot\hat{\bb{z}},
\end{align}
where $\bb{\sigma}$ and $\bb{\tau}$ are the vectors of Pauli matrices representing the spin and AFM sublattice degree of freedom, respectively; $\sigma_0$ is the identity matrix; $\bm{n}$ is the staggered order parameter (the normalized N\'{e}el vector); and $\alpha_R$ is the strength of the Rashba SOC. The kinetic energy of the itinerant electrons is $\gamma_k=a^2 t(k^2-k^2_0)$, where $t$ is the nearest-neighbor hopping energy and $k_0=2/a$, with $a$ being the lattice constant.

The band structure of the AFM-Rashba Hamiltonian of Eq.~(\ref{hamiltonian-AFM}) is, in general, anisotropic. It is convenient to parameterize the four spectral branches as follows:
\be
\label{energy-AFM}
E^{s,\eta}_k=\eta \sqrt{\gamma^2_k+J^2_{\mathrm{sd}}+\alpha^2_R k^2+2 s \alpha_R k \xi_k},
\e
where $s, \eta=\pm 1$ are the spin chirality and electron/hole band indices, respectively. We also introduce $\xi_k=\lt(\gamma^2_k+J^2_{\mathrm{sd}} \cos^2\phi\rt)^{1/2}$ with the in-plane wave vector $\bm{k}$ parameterized by the angle $\phi$, such that $\bm{k}=k(\cos\phi,\sin\phi,0)$.

\begin{figure}[t]
\includegraphics[width=8cm]{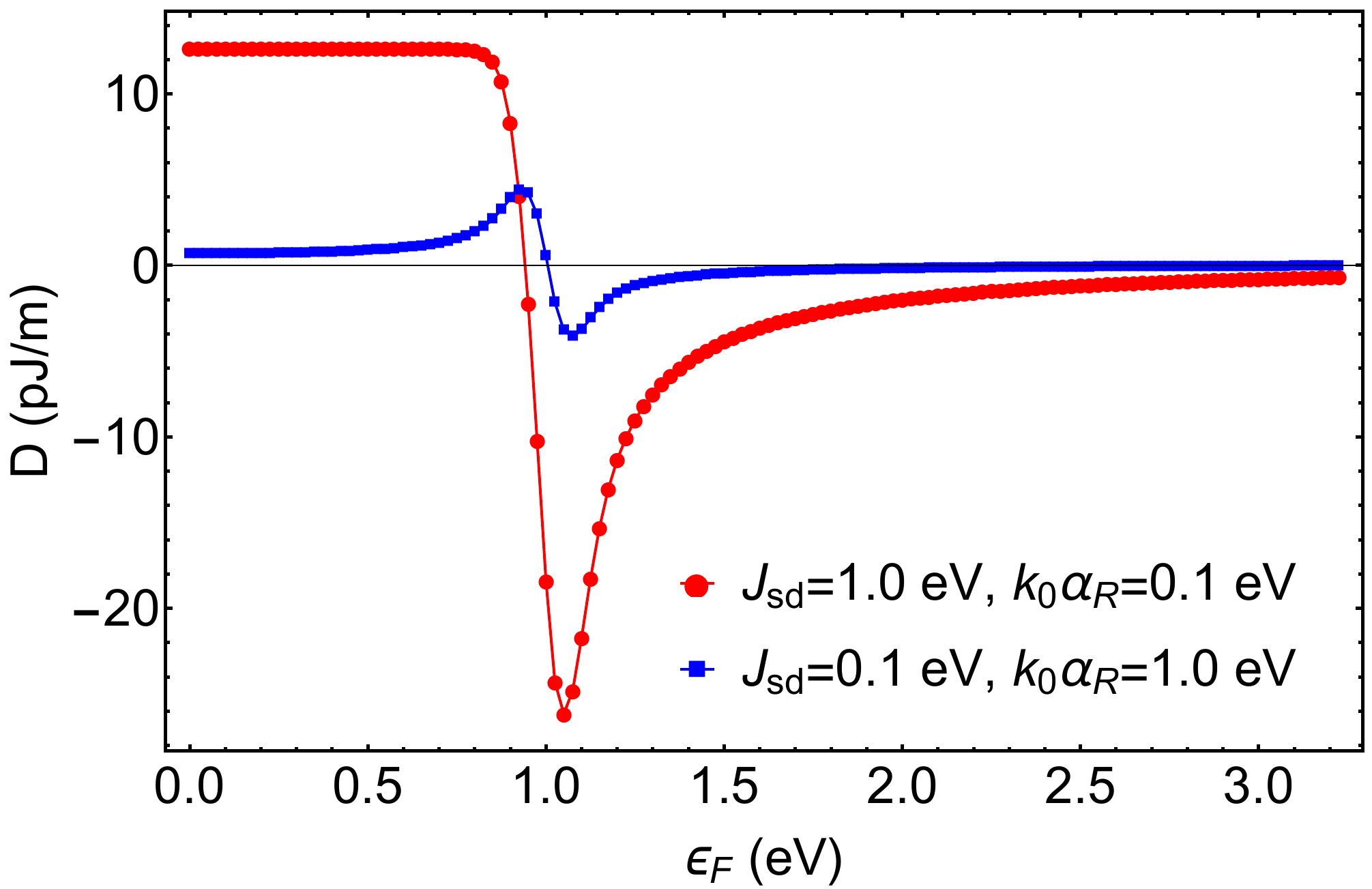}
\includegraphics[width=8cm]{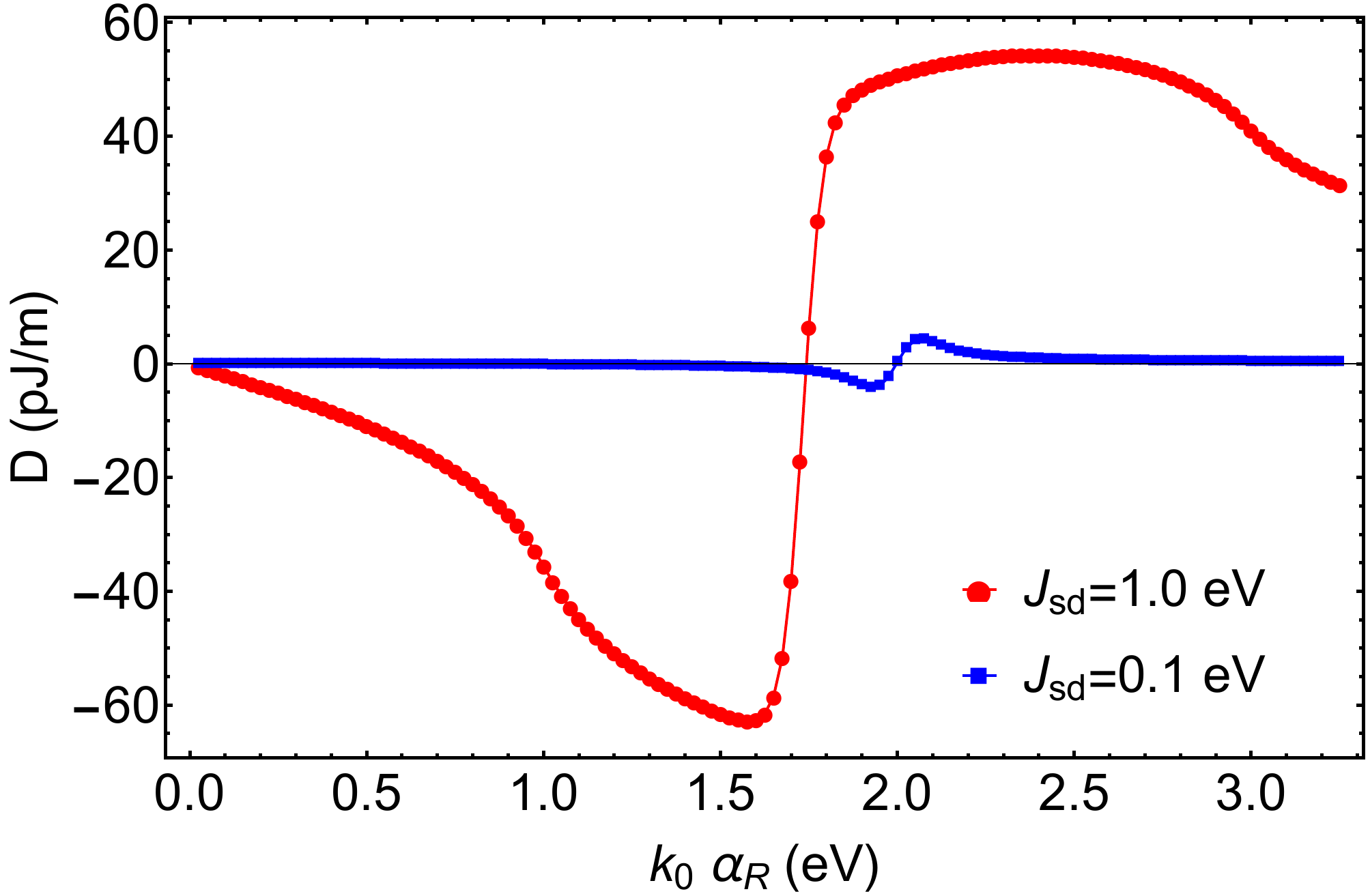}
\caption{Numerical results for the DM interaction as a function of the Fermi energy (top) and of the SOC strength (bottom) for different values of the \textit{s-d} exchange interaction. We set $\epsilon_F=2$ eV and $t=3$ eV in all cases.}
\label{fig:results}
\end{figure}

Let us now analyze the expressions of Eqs.~(\ref{DMI-formula22}) and (\ref{stiffness}) for the AFM-Rashba Hamiltonian of Eq.~(\ref{hamiltonian-AFM}) in the limit of weak spin-orbit interaction, $\{k_0 \alpha_R, J_\mathrm{sd}\} \ll \min\{t, \epsilon_F\}$, and at zero temperature. Using the relation $\Gamma_a=\tau_z\sigma_a$, we obtain the relation
\begin{align}
\label{DMI-AFM-weakSO}
D = -\left(\frac{A k^2_0}{4 t}\right)\alpha_R ,
\end{align}
where the stiffness parameter is
\be
\label{A-AFM-weakSO}
A = \frac{t J^2_{\mathrm{sd}}}{2 \pi \epsilon^2_F}
\bc
2-\epsilon^2_F/16 t^2,& \epsilon_F < 4t, \\
1,& \epsilon_F > 4t,
\ec
\e
which is manifestly independent of the SOC strength in the limit of weak spin-orbit interaction. Here, the Fermi energy $\epsilon_F$ is measured with respect to the center of the energy gap [see Figs.~\ref{fig:setup}(b) and (c)]. In this regime, the energy gaps at $k=0$ and $k=\pm k_0$ are equal to $8t$ and $2J_{\mathrm{sd}}$, respectively.

For the case in which the Fermi energy lies within the band gap, i.\,e., for $|\epsilon_F| < J_{\mathrm{sd}}$,  and in the limit of $k_0 \alpha_R \ll J_{\mathrm{sd}} \ll t$, we obtain
\begin{align}
\label{DMI-AFM3}
D =\left(\frac{k^2_0}{2\pi}\right) \alpha_R,
\end{align}
which is of the opposite sign compared with the metallic regime of Eqs.~(\ref{DMI-AFM-weakSO}) and (\ref{A-AFM-weakSO}).

It is instructive to compare the results of Eqs.~(\ref{DMI-AFM-weakSO})-(\ref{DMI-AFM3}) with those for a Rashba-FM system \cite{RashbaFM-DMI}. If both FM chiral bands are occupied ($\epsilon_F> J_{\mathrm{sd}}$), the interfacial DM interaction in the Rashba-FM model vanishes at zero temperature due to the exact cancellation between the Fermi surface and Fermi sea contributions \cite{RashbaFM-DMI}. However, such a cancellation is absent in the AFM-Rashba model, giving rise to a finite result for Eq.~(\ref{DMI-AFM-weakSO}), by virtue of an additional contribution from the valence bands. For $\epsilon_F < J_{\mathrm{sd}}$, i.\,e., when only the lowest FM chiral band is occupied, the DM interaction in the weak SOC limit and at zero temperature is finite and linearly proportional to $A$ \cite{iDMI7, Mokrousov, Tatara,RashbaFM-DMI}.

On the other hand, in the Dirac model of an FM/topological-insulator bilayer characterized by the Hamiltonian of the Rashba SOC symmetry, the DM interaction vanishes inside the gap but remains finite outside the gap, even at zero temperature\cite{iDMI-theory-TI1,iDMI-theory-TI2}. Thus, we conclude that the DM interaction exhibits a qualitatively different behavior in antiferromagnets compared with that in ferromagnets.

In Fig.~\ref{fig:results}, we illustrate the behavior of the DM interaction on the basis of a numerical analysis of Eq.~(\ref{DMI-formula2}) beyond the weak SOC regime of Eqs.~(\ref{DMI-AFM-weakSO}-\ref{DMI-AFM3}), where typical material parameters are assumed: $t=3$~eV, $\epsilon_F=2$~eV, $J_{\mathrm{sd}}=1$~eV, $a=0.4$~nm and $k_0\alpha_R=0.1$~eV \cite{AFM-Hamiltonian1}. The results indicate a rather large interfacial DM interaction, with $D\simeq -1.6$~pJ/m. Moreover, when the Fermi energy lies within the gap, i.\,e., $|\epsilon_F| < J_{\mathrm{sd}}$, we find an order-of-magnitude enhancement of the DM interaction with the opposite sign, $D\simeq 12.7$~pJ/m. Importantly, the estimated strength of the DM interaction is at least an order of magnitude larger than that for a Rashba-FM layer with the same parameters \cite{RashbaFM-DMI}.

For small Fermi energies, the DM interaction is positive and almost independent of the Fermi energy (see the top panel in Fig.~\ref{fig:results}), as might be expected from Eq.~(\ref{DMI-AFM3}). In the metallic regime (i.\,e., for Fermi energies well above the gap), the DM interaction is negative, and its strength decreases in proportion to $\epsilon_F^{-2}$, in good agreement with Eqs.~(\ref{DMI-AFM-weakSO}) and (\ref{A-AFM-weakSO}). The sign inversion of the DM interaction is rather sharp and occurs at $\epsilon_F=\sqrt{J^2_{\mathrm{sd}}+\alpha^2_R k^2_0}$. The bottom panel of Fig.~\ref{fig:results} also confirms that the strength of the DM interaction is linearly proportional to $\alpha_R$ in the weak SOC regime, in agreement with Eq.~(\ref{DMI-AFM-weakSO}). Thus, we conclude that the DM interaction in an AFM material may vary by orders of magnitude depending on the material parameters, as illustrated in Fig.~\ref{fig:results}.

Our calculations also show that, unlike in FM-Rashba systems \cite{RashbaFM-DMI}, the temperature dependence of the DM interaction in the AFM-Rashba model of Eq.~(\ref{hamiltonian-AFM}) is weak due to a large contribution to the DM interaction from the valence bands (not shown). As we have already discussed, this is correct if the system temperature is much less than the N\'{e}el temperature, i.\,e., in a regime in which the spin fluctuations of the AFM layer are suppressed.

Controlling the DM interaction is essential for engineering chiral magnetic structures. If the DM interaction parameter exceeds a certain critical value, which is determined by the Heisenberg exchange interaction and the uniaxial anisotropy, then the ground state changes from a collinear configuration to either a helimagnetic state or a Skyrmion lattice. A weaker DM interaction enables the stabilization of isolated Skyrmions in a metastable state \cite{Alireza-skyrmion,Yudin}. In chiral magnets, the sign of the DM interaction determines the direction, or handedness, of spin rotation.

The asymmetry of the spin-wave dispersion in spin-polarized electron energy-loss spectroscopy and Brillouin light scattering is a measure of the DM interaction in FM systems \cite{iDMI-exp0,iDMI-exp2,AFM-iDMI-exp1}. Although the interfacial DM interactions in a few FM/HM bilayers have been experimentally studied in recent years \cite{iDMI-exp1,iDMI-exp2,iDMI-exp3,iDMI-exp4,iDMI-exp5,iDMI-exp6,iDMI-exp7,iDMI-exp8,iDMI-exp9,iDMI-exp10,iDMI-exp11,iDMI-exp12,iDMI-exp13,AFM-iDMI-exp1}, we are not aware of similar measurements in AFM heterostructures. However, the first observation of a bulk DM interaction in an AFM system, namely, the noncentrosymmetric $\alpha-\mathrm{Cu_2V_2O_7}$, has recently been reported in Ref. \cite{AFMDMI-exp} based on inelastic neutron scattering. Very recently, a large DM interaction in Fe/Ir bilayers on Rh(001) has been predicted on the basis of {\em ab initio} calculations \cite{AFM-iDMI-DFT-2}. We hope that our work will stimulate new experiments and {\em ab initio} works on such AFM heterostructures.

In summary, we have computed the DM interaction, both analytically and numerically, using an effective model for AFM layers with interfacial Rashba SOC.
In the AFM-Rashba model, the induced interfacial DM interaction appears as a Lifshitz-type invariant term. Our results show that both the sign and the strength of the DM interaction may be tuned by modifying the electron density, e.\,g., by applying a gate voltage or through doping. This tunability implies that a rich variety of chiral magnetic structures can emerge in layered AFM/HM systems with different interfacial charge densities, \textit{s-d} exchange interactions, and SOC interactions.

\section*{Acknowledgments}
The research leading to these results was supported by the European Research Council via Advanced Grant No. 669442, ``Insulatronics," and by the Research Council of Norway through its Centres of Excellence funding scheme, Project No. 262633, ``QuSpin." We also acknowledge the support received from the Dutch Science Foundation, NWO/FOM 13PR3118; the European Commission; and the Russian Science Foundation under Project No. 17-12-01359. R. A. D. is part of the D-ITP consortium, a program of the Netherlands Organisation for Scientific Research (NWO) that is funded by the Dutch Ministry of Education, Culture and Science (OCW). M.T. acknowledges support from an ITMO visiting professor fellowship.


\begin{thebibliography}{9}
\bibitem {kzero}
In general, the effective action has two extra terms related to the spin susceptibilities with a long wavelength nature, i. e., $\bm{k}=0$, which have no contributions in the exchange stiffness and DM interaction:
$\Delta \mathcal{S}^{\mathrm{eff}}_{0}[\bm{n}]=\hbar\beta\left(\delta n^a \chi^{a}+\sum_{l}\delta n^a_{\nu_l} \Pi^{ab}_{i\nu_l} \delta n^b_{-\nu_l}\right)$. The static spin susceptibility related to the first-order term in $\delta n^a$ is $\chi^{a}=(J_{\mathrm{sd}}/\hbar\beta)\sum_{\bm{q},n}\mathrm{Tr}[\Gamma_a G^0_{\bm{q},i\omega_{n}}]$, and the dynamic spin susceptibility related to the second-order term in $\delta n^a$ is
$\Pi^{ab}_{i\nu_l}=\sum_{\bm{q},n}\big\{-(J_{\mathrm{sd}}/2\hbar\beta)\mathrm{Tr}[\Gamma_x G^0_{\bm{q},i\omega_{n}}]\delta_{ab}+(J^2_{\mathrm{sd}}/2\hbar\beta)\mathrm{Tr}[\Gamma_a G^0_{\bm{q},i\omega_{n}}\Gamma_b G^0_{\bm{q},i\omega_{n}+i\nu_l}]\big\}$, where $\delta_{ab}$ is the Kronecker delta function.
\end{thebibliography}
\end{document}